\documentclass[a4paper,10pt]{article}
\usepackage{amsmath,amssymb}

\title{{\bf Some properties of partial fidelities}}

\author{Alexey E. Rastegin}
\date{\small Department of Theoretical Physics, Irkutsk State
University, Gagarin Bv. 20, Irkutsk 664003, Russia}

\begin{document}
\maketitle
\begin{abstract}
Basic properties of Uhlmann's partial fidelities are discussed.
Statistical interpretation in terms of POVM measurements is
established. Multiplicativity properties are considered. The
relationship between partial fidelities and partitioned trace
distances is derived. As it is shown, the partial fidelities
cannot decrease under unistochastic quantum operations. Thus, the
partial fidelities have good properties in the sense of their use
as distinguishability measures.

\vspace{5mm}
\end{abstract}

\section{Introduction}

The concept of fidelity provides a very useful measure of
distinguishability \cite{uhlmann76,uhlmann83,jozsa94}. It is well
known that all the devices for quantum information processing are
inevitably exposed to noise \cite{nielsen}. Therefore, the pure
states used by us will eventually evolve to mixed states. Thus,
good distance measures between mixed quantum states are needed for
comparison of real and ideal processes \cite{nielsen05}. For
example, different measures of distinguishability were utilized
for study of probability simplex \cite{braunstein}, quantum
entanglement \cite{bengtsson}, approximate broadcasting
\cite{holevo} and quantum circuit complexity \cite{nielsen08}. If
states are pure then comparison of them is not difficult. In
contrast, there exist many ways to compare two mixed quantum
states.

Both the fidelity and the trace distance are extensively utilized
as measures of distinguishability \cite{nielsen,fuchs}. However,
these two measures are not able to pose the problem of state
closeness in all respects. For instance, the equality of
fidelities for two pairs of density operators does not imply their
unitary equivalence \cite{uhlmann00}. The justified answer is
provided by use of the partial fidelities \cite{uhlmann00}. In the
literature, other fidelity measures between quantum states have
been discussed \cite{foster,uhlmann09}. There are also some
closely related measures such as the Bures distance
\cite{braunstein,bengtsson}, the angle \cite{nielsen,rast031}, the
Monge distance \cite{kzws01} and the sine distance \cite{rast06}.

In principle, reasons for use of one or another distance measure
are especially based on main properties of the measure. These
properties are usually connected with the measurements, behaviour
under quantum operations and the convexity (concavity) in inputs.
For convenience, they must provide useful mathematical tools for
the work. In addition, the properties should have clear physical
interpretation. In the present paper, we discuss the partial
fidelities introduced by Uhlmann. In Ref. \cite{uhlmann00} Uhlmann
paid key attention to the equivalence of pairs of mixed states
under invertible transformations of the state space. The aim of
the present work is to analyze partial fidelities in those
respects that are not considered in Ref. \cite{uhlmann00}. In
particular, quantum--classical relations and behaviour under
quantum operations are examined.

\section{Definitions and background}

In this section, we shall briefly review necessary facts from
linear algebra. Then the basic definitions will be given. Let
${\cal{H}}$ be $d$-dimensional Hilbert space. By
${\rm{supp}}({\mathsf{X}})$ we denote the support of operator
${\mathsf{X}}$. Putting an inner product function for operators on
${\cal{H}}$, we define
\begin{equation}
\langle{\mathsf{X}},{\mathsf{Y}}\rangle_{\rm{hs}}
:={\rm{tr}}({\mathsf{X}}^{\dagger}{\mathsf{Y}})
\ . \label{hsdef}
\end{equation}
This is the Hilbert--Schmidt inner product of operators
${\mathsf{X}}$ and ${\mathsf{Y}}$ \cite{watrous1}. For any
operator ${\mathsf{X}}$ on ${\cal{H}}$ the operator
${\mathsf{X}}^{\dagger}{\mathsf{X}}$ is positive, that is
$\langle\psi|{\mathsf{X}}^{\dagger}{\mathsf{X}}|\psi\rangle\geq0$
for all $|\psi\rangle\in{\cal{H}}$. The operator $|{\mathsf{X}}|$
is defined as unique positive square root of
${\mathsf{X}}^{\dagger}{\mathsf{X}}$. The eigenvalues of operator
$|{\mathsf{X}}|$ counted with their multiplicities are named {\it
singular values} of operator ${\mathsf{X}}$
\cite{watrous1,bhatia}. As a rule, we will deal with eigenvalues
of Hermitian operator. So these eigenvalues are real numbers. In
the text, the eigenvalues of operator ${\mathsf{X}}$ are assumed
to be indexed in {\it decreasing order}, that is
\begin{equation}
{\rm{spec}}({\mathsf{X}})=
\{\lambda_1({\mathsf{X}})\geq\lambda_2({\mathsf{X}})\geq\cdots\geq\lambda_d({\mathsf{X}})\}
\ . \label{xspec}
\end{equation}
The singular values of ${\mathsf{X}}$ are then denoted by
$s_1({\mathsf{X}})\geq\cdots\geq s_d({\mathsf{X}})$. So the $k$-fidelity between two mixed states is defined
as follows \cite{uhlmann00}.

{\bf Definition 1.} {\it The $k$-th partial fidelity between
density operators $\rho$ and $\omega$ is the sum of $(d-k)$
smallest singular values of $\sqrt{\rho}\sqrt{\omega}$, that is}
\begin{equation}
F_k(\rho,\omega):=\sum\nolimits_{j>k}   s_j(\sqrt{\rho}\sqrt{\omega})
\ . \label{defin1}
\end{equation}

Uhlmann showed \cite{uhlmann00} that the partial fidelities are
symmetric, unitarily invariant and jointly concave. The proof of
joint concavity is based on the following result. We consider all
the pairs $\{{\mathsf{X}},{\mathsf{Y}}\}$ of positive operators
that satisfy
\begin{eqnarray}
 & {\mathsf{X}}{\mathsf{Y}}{\mathsf{X}}={\mathsf{X}}
\ , \label{xpair} \\
 & {\mathsf{Y}}{\mathsf{X}}{\mathsf{Y}}={\mathsf{Y}}
\ . \label{ypair}
\end{eqnarray}
It have been shown that ${\rm{rank}}({\mathsf{X}})={\rm{rank}}({\mathsf{Y}})={\rm{rank}}({\mathsf{XY}})$
\cite{uhlmann00}. Then the $k$-th partial fidelity can be expressed as \cite{uhlmann00}
\begin{equation}
F_k(\rho,\omega)=\frac{1}{2}
\>\inf\bigl\{{\rm{tr}}(\rho{\mathsf{X}})+{\rm{tr}}(\omega{\mathsf{Y}})\bigr\}
\ , \label{uhlres}
\end{equation}
where the infimum is taken over all the above pairs of rank
$(d-k)$. In general, the sign "inf" cannot be replaced by the sign
"min" in Eq. (\ref{uhlres}). In the case $k=0$, Definition 1 is
reduced to the well-known formula for usual fidelity. Indeed, we
obtain
\begin{equation}
F_0(\rho,\omega)=\sum\nolimits_{j=1}^d s_j(\sqrt{\rho}\sqrt{\omega})
\equiv{\rm{tr}}|\sqrt{\rho}\sqrt{\omega}|
\ , \label{defin11}
\end{equation}
and the latter is the fidelity \cite{nielsen,caves}. The square of
this value was introduced by Uhlmann as "transition probability"
between two states \cite{uhlmann76}. Note that Jozsa
\cite{jozsa94} used the word "fidelity" just for Uhlmann's
transition probability. This is more convenient in some tasks of
quantum information processing, for instance in quantum cloning
\cite{rast031,rast032}. However, in the present paper we shall use
the term "fidelity" for the expression in Eq. (\ref{defin11}). In
the physical literature the basic properties of fidelity are
usually given by staying in finite dimensions. But they hold in
much more generality \cite{alulm00}.

Distinguishability measures between quantum states are closely
related to the classical distinguishability measures. Let
$\{p_i\}$ and $\{q_i\}$ be two probability distributions over the
same index set. The fidelity between these distributions is
defined by \cite{fuchs,caves}
\begin{equation}
{\cal{F}}(p_i,q_i):=\sum\nolimits_i \sqrt{p_i q_i} \ .
\label{fidcl}
\end{equation}
There is a direct analogy between the right-hand side of Eq.
(\ref{defin11}) and the right-hand side of Eq.(\ref{fidcl}). Let
integer $r$ denote the cardinality of the sets $\{p_i\}$ and
$\{q_i\}$. In the same manner, the classical $k$-fidelity can be
introduced as the sum of $(r-k)$ smallest numbers $\sqrt{p_iq_i}$.

{\bf Definition 2.} {\it The $k$-th partial fidelity between
probability distributions $\{p_i\}$ and $\{q_i\}$ is defined by}
\begin{equation}
{\cal{F}}_k(p_i,q_i):=\sum\nolimits_{i>k} \sqrt{p_i q_i}^{\>\downarrow} \ .
\label{partfidcl}
\end{equation}
{\it where values $\sqrt{p_i q_i}^{\>\downarrow}$ are indexed in the decreasing order.}

Like the standard fidelity, the quantum $k$-fidelity are closely
related to the classical $k$-fidelity for commuting density
operators $\rho$ and $\omega$. In this simple case the two density
operators are diagonal in the same orthonormal basis
$\{|i\rangle\}$, namely
\begin{eqnarray}
& \rho=\sum\nolimits_i \mu_i
\>|i\rangle\langle i| \ , \label{rhocomm} \\
& \omega=\sum\nolimits_i \nu_i
\>|i\rangle\langle i| \ . \label{varcomm}
\end{eqnarray}
The operator
$\sqrt{\rho}\sqrt{\omega}=\sum\nolimits_i\sqrt{\mu_i\nu_i}\>|i\rangle\langle
i|$ has singular values $\sqrt{\mu_i\nu_i}^{\>\downarrow}$. Due to
Definitions 1 and 2, we at once obtain
\begin{equation}
F_k(\rho,\omega)={\cal{F}}_k(\mu_i,\nu_i)
\ . \label{comm}
\end{equation}
In the next section, a connection between quantum and classical will be given in terms of quantum measurements.

Now we write down some simple properties of $k$-fidelity. These
facts are based on the following easy reason. Let $\{a_i\}$ be a
set of $r$ positive numbers arranged in the decreasing order. We
then define two quantity
\begin{eqnarray}
 & A_k := \sum\nolimits_{1\leq i\leq k} a_i
\ , \label{ax0} \\
 & B_k := \sum\nolimits_{k+1\leq i \leq r}  a_i
\ . \label{bx0}
\end{eqnarray}
When the index $i$ ranges the empty set, any sum is assumed to be zero.

{\bf Lemma 3.} {\it For all $k=0,1,\ldots,r$, there holds}
\begin{eqnarray}
& (k+1) A_k \geq k A_{k+1}
\ , \label{ax1} \\
& (r-k-1) B_k \geq (r-k) B_{k+1}
\ . \label{bx1}
\end{eqnarray}

Iterating Eq. (\ref{bx1}), we see that both the classical and quantum partial fidelities obey
\begin{equation}
{\cal{F}}_k(p_i,q_i) \leq\frac{r-k}{r}\>{\cal{F}}_0(p_i,q_i)
\ , \ \ F_k(\rho,\omega) \leq\frac{d-k}{d}\> F_0(\rho,\omega)
\ . \label{fkinq}
\end{equation}
Since $0$-fidelity does not exceed one, we have
${\cal{F}}_k(p_i,q_i)\leq(r-k)/r$ and
$F_k(\rho,\omega)\leq(d-k)/d$. The last two inequalities are
saturated if and only if $\{p_i\}=\{q_i\}$ and $\rho=\omega$.
Further, ${\cal{F}}_k(p_i,q_i)=0$ for all $k$, if and only if the
distributions do not overlap at all. Similarly,
$F_k(\rho,\omega)=0$ for all $k$, if and only if the subspaces
${\rm{supp}}(\rho)$ and ${\rm{supp}}(\omega)$ do not intersect.

\section{Quantum-classical relations via measurement}

In this section, we provide a statistical interpretation for
partial fidelities between density operators. Reasons are
formulated in terms of probabilities generated by a quantum
measurement. A general form of quantum measurement is described by
so-called "positive operator-valued measure" (POVM). Recall that
POVM $\{{\mathsf{M}}_m\}$ is a set of positive operators
${\mathsf{M}}_m$ satisfying the completeness relation
\begin{equation}
\sum\nolimits_m {\mathsf{M}}_m={\mathbf{1}}
\ , \label{povmdef}
\end{equation}
where ${\mathbf{1}}$ is the identity operator on the space
${\cal{H}}$ \cite{peres}. For two density operators, the traces
${\rm tr}({\mathsf{M}}_m\rho)\equiv p_m$ and ${\rm
tr}({\mathsf{M}}_m\omega)\equiv q_m$ are the probabilities of
obtaining a measurement result labeled by $m$. Measuring concrete
observable, we are concerned with projective measurement described
by so-called "projection valued measure" (PVM) \cite{peres}.

The explicit statistical interpretation for the fidelity was
derived in Ref. \cite{caves}. The authors of Ref. \cite{caves}
showed that the fidelity satisfies
\begin{equation}
F(\rho,\omega)=\min {\cal{F}}(p_m,q_m)
\ , \label{relfid0}
\end{equation}
where the minimization is over all POVMs. So, the quantum
fidelity is achievable lower bound for the classical fidelity
generated by measurement. By relevant modification, this property
can be extended to partial fidelities. Before the assertion, we
recall a useful fact from matrix theory. In the seminal paper
\cite{kyfan} Ky Fan obtained a lot of connected results with
respect to extremal properties of eigenvalues. One of his
formulations is now known as Ky Fan's maximum principle
\cite{bhatia}. This principle is one of important results in
majorization theory \cite{vidal}. In our notation, the second part of Theorem 1 of
Ref. \cite{kyfan} says that for any Hermitian operator ${\mathsf{X}}$
\begin{equation}
\sum\nolimits_{j>k} \lambda_j({\mathsf{X}})=
\min \{{\rm tr}({\mathsf{\Pi}}{\mathsf{X}}):\>{\rm{rank}}({\mathsf{\Pi}})=d-k\}
\ , \label{kfminpr}
\end{equation}
where the minimum is taken over all projectors ${\mathsf{\Pi}}$ of
rank $(d-k)$. For positive ${\mathsf{X}}$, the condition can be
replaced by ${\rm{rank}}({\mathsf{\Pi}})\geq d-k$. (Recall that we
assume the decreasing order of eigenvalues.) The minimum is
obviously reached by the projector ${\mathsf{\Pi}}_{\rm{min}}$
onto the subspace corresponding to $(d-k)$ smallest eigenvalues of
operator ${\mathsf{X}}$. Modifying the reasons given in Ref.
\cite{kyfan}, we can state that for positive ${\mathsf{X}}$,
\begin{equation}
\sum\nolimits_{j>k} \lambda_j({\mathsf{X}})=
\min \{{\rm tr}({\mathsf{T}}{\mathsf{X}}):
\>{\mathbf{0}}\leq{\mathsf{T}}\leq{\mathbf{1}},\>{\rm tr}({\mathsf{T}})\geq d-k\}
\ , \label{kfminpr2}
\end{equation}
where the minimization is over those positive operators
${\mathsf{T}}$ that obey ${\mathsf{T}}\leq{\mathbf{1}}$ and ${\rm
tr}({\mathsf{T}})\geq d-k$. We do not enter into details here.

{\bf Theorem 4.} {\it If all the POVM elements satisfy
${\rm{tr}}({\mathsf{M}}_m)\geq 1$, then for arbitrary two density
operators $\rho$ and $\omega$ there holds}
\begin{equation}
F_k(\rho,\omega)\leq {\cal{F}}_k(p_m,q_m)
\ . \label{relfid2}
\end{equation}

{\bf Proof.} By the polar decomposition \cite{nielsen}, there is
$|\sqrt{\rho}\sqrt{\omega}|=\sqrt{\rho}\sqrt{\omega}\>{\mathsf{V}}$
for some unitary ${\mathsf{V}}$. It is known that operators
${\mathsf{X}}{\mathsf{Y}}$ and ${\mathsf{Y}}{\mathsf{X}}$ have the
same eigenvalues (see, for instance, Exercise A6.5 in Ref.
\cite{nielsen}). Therefore, we have
\begin{equation}
{\rm{spec}}(|\sqrt{\rho}\sqrt{\omega}|)={\rm{spec}}(\sqrt{\omega}\>{\mathsf{V}}\sqrt{\rho})
\ . \label{specrhom}
\end{equation}
For given POVM $\{{\mathsf{M}}_m\}$ and two density operators
$\rho$ and $\omega$, we rearrange POVM elements with respect to
the decreasing order of numbers $\sqrt{p_m q_m}^{\>\downarrow}$.
Let us define new positive operator
\begin{equation}
{\mathsf{T}}_M= \sum\nolimits_{m>k} {\mathsf{M}}_m^{\>\downarrow}
\ . \label{lamm}
\end{equation}
Due to the completeness relation, one obeys
${\mathsf{T}}_M\leq{\mathbf{1}}$. Each of $(d-k)$ terms in Eq.
(\ref{lamm}) has trace exceeding one, whence
${\rm{tr}}({\mathsf{T}}_M)\geq d-k$. Due to Eqs. (\ref{kfminpr2})
and (\ref{specrhom}), we can write
\begin{equation}
F_k(\rho,\omega)\leq {\rm{tr}}({\mathsf{T}}_M\sqrt{\omega}\>{\mathsf{V}}\sqrt{\rho})
=\sum\nolimits_{m>k} {\rm{tr}}(\sqrt{\rho}\>{\mathsf{M}}_m^{\>\downarrow}
\sqrt{\omega}\>{\mathsf{V}})
\ . \label{proff31}
\end{equation}
Let us define operators
${\mathsf{A}}_m=({\mathsf{M}}_m^{\>\downarrow})^{1/2}{\rho}^{1/2}$
and
${\mathsf{B}}_m=({\mathsf{M}}_m^{\>\downarrow})^{1/2}{\omega}^{1/2}\>{\mathsf{V}}$.
In Eq. (\ref{proff31}) each term of the sum is the
Hilbert--Schmidt inner product
$\langle{\mathsf{A}}_m,{\mathsf{B}}_m\rangle_{\rm{hs}}$ satisfying
\begin{equation}
\left|\langle{\mathsf{A}}_m,{\mathsf{B}}_m\rangle_{\rm{hs}}\right| \leq
\langle{\mathsf{A}}_m,{\mathsf{A}}_m\rangle_{\rm{hs}}^{\>1/2}
\langle{\mathsf{B}}_m,{\mathsf{B}}_m\rangle_{\rm{hs}}^{\>1/2}
\ . \label{abhs}
\end{equation}
Using the cyclic property of the trace, we have
\begin{eqnarray}
&   \langle{\mathsf{A}}_m,{\mathsf{A}}_m\rangle_{\rm{hs}}=
{\rm{tr}}({\mathsf{M}}_m^{\>\downarrow}\rho)
\ , \label{amam} \\
&   \langle{\mathsf{B}}_m,{\mathsf{B}}_m\rangle_{\rm{hs}}=
{\rm{tr}}({\mathsf{M}}_m^{\>\downarrow}\omega)
\ . \label{bmbm}
\end{eqnarray}
Together with Eq. (\ref{abhs}), the last equalities imply that
\begin{equation}
\sum\nolimits_{m>k}\left| {\rm{tr}}(\sqrt{\rho}\>{\mathsf{M}}_m^{\>\downarrow}
\sqrt{\omega}\>{\mathsf{V}}) \right| \leq \sum\nolimits_{m>k}\sqrt{p_m q_m}^{\>\downarrow}
\ . \label{trabhs}
\end{equation}
Inequalities (\ref{proff31}) and (\ref{trabhs}) provide Eq.
(\ref{relfid2}) for every POVM whose elements fulfill the
precondition ${\rm{tr}}({\mathsf{M}}_m)\geq 1$. $\blacksquare$

For the $0$-fidelity, the lower bound ${\cal{F}}_0(p_m,q_m)$ can
be reached by some POVM's \cite{nielsen,caves}. For invertible
$\rho$, the optimal POVM is formed by one-rank projectors
${\mathsf{P}}_m=|m\rangle\langle m|$, where $|m\rangle$ is
eigenstate of positive operator
${\mathsf{R}}=\rho^{-1/2}|\sqrt{\rho}\sqrt{\omega}|\rho^{-1/2}$.
Here we have a proportionality
\begin{equation}
\lambda_m({\mathsf{R}})\>{\mathsf{P}}_m\sqrt{\rho}=
{\mathsf{P}}_m\sqrt{\omega}\>{\mathsf{V}}
\ . \label{propor}
\end{equation}
Therefore, the Schwarz inequality (\ref{abhs}) is fulfilled with
equality for all $m$. However, this is not sufficient for partial
fidelity of order $k>0$. To saturate inequality (\ref{relfid2}),
we must also provide the equality in Eq. (\ref{proff31}). But the
operator
\begin{equation}
{\mathsf{T}}_P= \sum\nolimits_{m>k} {\mathsf{P}}_m^{\>\downarrow}
\ . \label{lapp}
\end{equation}
is not the projector ${\mathsf{\Pi}}_{\rm{min}}$ with necessity.
When ${\mathsf{T}}_P={\mathsf{\Pi}}_{\rm{min}}$ for all $k$, the
operators ${\mathsf{R}}$ and $|\sqrt{\rho}\sqrt{\omega}|$ are
simultaneously diagonalizable. Thus, the operators ${\mathsf{R}}$
and $|\sqrt{\rho}\sqrt{\omega}|^2=\sqrt{\rho}\>\omega\sqrt{\rho}$
are commuting, whence
\begin{equation}
\rho\>\omega|\sqrt{\rho}\sqrt{\omega}|=|\sqrt{\rho}\sqrt{\omega}|\rho\>\omega
\ . \label{commut}
\end{equation}
So we have arrived at a conclusion. If the inequality
(\ref{relfid2}) can be saturated for all $k=0,\ldots,d-1$, then
the operators $\rho\>\omega$ and $|\sqrt{\rho}\sqrt{\omega}|$ are
commuting.

It should be pointed out that the precondition
${\rm{tr}}({\mathsf{M}}_m)\geq 1$ of Theorem 4 is necessary.
Indeed, for POVM with several elements of the form
${\mathsf{N}}_m=\epsilon\>|m\rangle\langle m|$ proper
fidelities ${\cal{F}}_k(p_m,q_m)$ can be made arbitrarily small
independently of $F_k(\rho,\omega)$. But such POVM's can unlikely
be interesting in the practice. For many applications of quantum
information we primarily deal with projective measurements
\cite{nielsen}. The statement of Theorem 4 holds for all the
projective measurements. From the operational point of view, the
attainability of the lower bound in Eq. (\ref{relfid2}) is hardly
essential. In quantum information processing, we usually have only
partial knowledge about quantum states. So the measurement
optimizing Eq. (\ref{relfid0}) cannot be found {\it a priori}. In
this situation the right-hand sides of Eqs. (\ref{relfid0}) and
(\ref{relfid2}) provide measurable estimates from above on the
quantum fidelities.

\section{Sub--multiplicativity and relation with trace distances}

It is usual in the study of quantum information that one deals
with composite systems \cite{nielsen,watrous1}. Here the
multiplicativity of fidelity makes this measure very convenient to
use. That is, if $\rho$ and $\omega$ are density operators on
${\cal{H}}_A$, $\Theta$ and $\Omega$ are density operators on
${\cal{H}}_E$, then
\begin{equation}
F_0(\rho\otimes\Theta,\omega\otimes\Omega)
=F_0(\rho,\omega)\>F_0(\Theta,\Omega)
\ . \label{mult}
\end{equation}
The authors of Ref. \cite{uhlmann09} introduced two quantity,
namely sub--fidelity and super--fidelity. These measure are easier
to experimental measurement than the fidelity. The sub--fidelity
and super--fidelity provide the lower and upper bounds on the
fidelity respectively. In terms of super--fidelity the strong
lower bound on the trace distance has been established
\cite{puchala}. The sub–-fidelity is sub–-multiplicative, the
super-–fidelity is super-–multiplicative \cite{uhlmann09}. We
shall now find analogous property for the partial fidelities.

There is simple property of singular values with respect to the
operation of tensor product \cite{watrous1}. Namely, the singular
values of tensor product of two operators are the products of all
pairs consisting of a singular value of first and a singular value
of  second operator. A justification is the following. Recall that
for arbitrary four operators of proper dimensionality there holds
\cite{watrous1}
\begin{equation}
({\mathsf{X}}\otimes\Theta)({\mathsf{Y}}\otimes\Omega)=({\mathsf{XY}})\otimes(\Theta\Omega)
\ . \label{proptens}
\end{equation}
Using this property and $\sqrt{{\mathsf{X}}\otimes\Theta}=\sqrt{\mathsf{X}}\otimes\sqrt{\Theta}$, we at once obtain
\begin{equation}
|{\mathsf{X}}\otimes\Theta|=\sqrt{({\mathsf{X}}\otimes\Theta)^{\dagger}({\mathsf{X}}\otimes\Theta)}=
\sqrt{({\mathsf{X}}^{\dagger}{\mathsf{X}})\otimes(\Theta^{\dagger}\Theta)}
=|{\mathsf{X}}|\otimes|\Theta|
\ . \label{tens}
\end{equation}
In the same manner, we also have
\begin{equation}
\sqrt{\rho\otimes\Theta}\>\sqrt{\omega\otimes\Omega}=
\sqrt{\rho}\sqrt{\omega}\otimes\sqrt{\Theta}\sqrt{\Omega}
\ . \label{sqsq}
\end{equation}
Let $|x\rangle$ be eigenvector of $|{\mathsf{X}}|$, and let
$|\theta\rangle$ be eigenvector of $|\Theta|$. By Eq.
(\ref{tens}), the product $|x\rangle\otimes|\theta\rangle$ is
eigenvector of $|{\mathsf{X}}\otimes\Theta|$. So we get the
following. If $d$ numbers $s_j({\mathsf{X}})$ are singular values
of operator ${\mathsf{X}}$, $N$ numbers $s_i(\Theta)$ are singular
values of operator $\Theta$, then $dN$ products
\begin{equation}
s_{ji}({\mathsf{X}}\otimes\Theta)=s_j({\mathsf{X}})\>s_i(\Theta)
\label{prod}
\end{equation}
give all the singular values of operator
${\mathsf{X}}\otimes\Theta$ in {\it a priori} unknown order. In
accordance with Definition 1, we have
\begin{equation}
F_{(d-k)}(\rho,\omega)\>F_{(N-L)}(\Theta,\Omega)=\sum\nolimits_{j=d-k+1}^{d}
s_j^{\downarrow}(\sqrt{\rho}\sqrt{\omega})
\>\sum\nolimits_{i=N-L+1}^{N} s_i^{\downarrow}(\sqrt{\Theta}\sqrt{\Omega})
\ . \label{dknl}
\end{equation}
Due to Eqs. (\ref{tens}), (\ref{sqsq}) and (\ref{prod}), the
right-hand side of Eq. (\ref{dknl}) keeps a sum of $kL$
singular values of operator
$\sqrt{\rho\otimes\Theta}\>\sqrt{\omega\otimes\Omega}$. The partial fidelity of order $(dN-kL)$ between
$\rho\otimes\Theta$ and $\omega\otimes\Omega$ does not exceed this sum. So, there holds
\begin{equation}
F_{(dN-kL)}(\rho\otimes\Theta,\omega\otimes\Omega)\leq
F_{(d-k)}(\rho,\omega)\>F_{(N-L)}(\Theta,\Omega)
\ . \label{submult}
\end{equation}
Therefore, the partial fidelities are sub--multiplicative in the
sense of Eq. (\ref{submult}). When $k=d$ and $L=N$, the right-hand
side of Eq. (\ref{dknl}) summarized all the singular values of
operator $\sqrt{\rho\otimes\Theta}\>\sqrt{\omega\otimes\Omega}$.
So, the inequality (\ref{submult}) is replaced by Eq.
(\ref{mult}). Thus, the multiplicative property of partial
fidelities is more complicated in character. In general, this is
not unexpected.

On a level with properties of some measure itself, the
relationships between different measures of distinguishability are
very interesting. It is well-known that the fidelity and the trace
distance are related by the inequalities \cite{graaf}
\begin{equation}
1-F(\rho,\omega)\leq D(\rho,\omega) \leq\sqrt{1-F(\rho,\omega)^2}
\ . \label{fidtr}
\end{equation}
Here the trace distance between density operators $\rho$ and $\omega$ is defined by
\begin{equation}
D(\rho,\omega):=\frac{1}{2}\>{\rm tr}|\rho-\omega|
\ . \label{tracedis}
\end{equation}
There is an alternative definition via extremal properties of
quantum operations \cite{rast07}. Therefore, these measures may be
considered to be equivalent for many applications
\cite{nielsen,watrous1}. A relative analog of the above
relationship can be obtained in terms of {\it partitioned trace
distances}. In Ref. \cite{rast091}, the present author imported
this family of new distances between mixed quantum states. For
$k=1,\ldots,d$, the Ky Fan $k$-norm of operator ${\mathsf{X}}$ is
defined by \cite{bhatia}
\begin{equation}
||{\mathsf{X}}||_{(k)}:=\sum\nolimits_{j=1}^{k}
s_j({\mathsf{X}}) \ ,
\label{fannorm}
\end{equation}
where singular values are assumed to be indexed in decreasing
order. The $k$-th partitioned trace distance between density
operator $\rho$ and $\omega$ is introduced as
\begin{equation}
D_k(\rho,\omega):=\frac{1}{2}\>||\rho-\omega||_{(k)}
\ . \label{defpard}
\end{equation}
The partitioned distances succeed many properties of the standard
trace distance, namely the unitary invariance, the strong
convexity and the bounds. In Ref. \cite{rast091}, the present
author also defined the $k$-th classical distance between two
probability distributions $\{p_i\}$ and $\{q_i\}$ as
\begin{equation}
{\cal{D}}_k(p_i,q_i):=\frac{1}{2}\>\sum\nolimits_{i=1}^{k} |p_i-q_i|^{\downarrow}
\ , \label{pcltracedis}
\end{equation}
where the arrows down indicate that the absolute values are put in
the decreasing order. Using the Ky Fan maximum principle, we can
prove that \cite{rast091}
\begin{equation}
D_k(\rho,\omega)=\max\{ {\cal{D}}_k(p_m,q_m):
\> {\rm{tr}}({\mathsf{M}}_m)\leq1 \}
\ , \label{relt0}
\end{equation}
where the maximum is taken over all POVMs $\{{\mathsf{M}}_m\}$
whose elements satisfy ${\rm{tr}}({\mathsf{M}}_m)\leq1$. The
maximum in Eq. (\ref{relt0}) is reached by one-rank PVM
$\{{\mathsf{Q}}_m\}$ associated with the Jordan decomposition of
difference $(\rho-\omega)$ \cite{rast091}. We rearrange elements
of this optimal PVM with respect to the decreasing orders of numbers
$|p_m-q_m|^{\downarrow}$. Let us define projector
${\mathsf{\Pi'}}$ of rank $(d-k)$ by
\begin{equation}
{\mathsf{\Pi'}}:=\sum\nolimits_{m>k} {\mathsf{Q}}_m^{\>\downarrow}
\ . \label{pisch}
\end{equation}
Putting ${\mathsf{X}}={\mathsf{\Pi'}}$ and
${\mathsf{Y}}={\mathsf{\Pi'}}$, we satisfy Eqs. (\ref{xpair}) and
(\ref{ypair}). Due to Eq. (\ref{uhlres}), we then have
\begin{eqnarray}
2\>F_k(\rho,\omega)&\leq & {\rm{tr}}({\mathsf{\Pi'}}\rho)+  {\rm{tr}}({\mathsf{\Pi'}}\omega)
\nonumber\\
&=&1-\sum_{m=1}^{k} p_m +1-\sum_{m=1}^{k} q_m
\nonumber\\
&\leq & 2- \sum\nolimits_{m=1}^{k} |p_m-q_m|^{\downarrow}
\ . \label{serineq}
\end{eqnarray}
By choice of optimal PVM, the
right-hand side of Eq. (\ref{serineq}) is equal to quantity
$(2-2\>D_k(\rho,\omega))$. Hence we obtain
\begin{equation}
F_k(\rho,\omega)+D_k(\rho,\omega)\leq 1
\ . \label{fkled}
\end{equation}
This inequality should not be confused with inequalities
(\ref{fidtr}) which contain $F_0(\rho,\omega)$ and
$D_d(\rho,\omega)$. Using Eq. (\ref{ax1}), we easily derive that
\begin{equation}
\frac{k}{r}\>{\cal{D}}_r(p_i,q_i) \leq {\cal{D}}_k(p_i,q_i)
\ , \ \ \frac{k}{d}\> D_d(\rho,\omega) \leq D_k(\rho,\omega)
\ . \label{dkinq}
\end{equation}
By Eq. (\ref{fkinq}), the $k$-th partial fidelity is bounded from
above. By Eq. (\ref{dkinq}), the $k$-th partitioned trace distance
is bounded from below. So the inequality (\ref{fkled}) is not
trivial, although it differs from inequalities (\ref{fidtr}) in
character. Thus, if the $k$-th partial fidelity between two states
is close to one then the $k$-th partitioned trace distance is
close to zero. In the mentioned sense, these measures of closeness
for quantum states can be regarded as equivalent.

\section{Monotonicity under unistochastic quantum operations}

One of the basic properties of the fidelity is that fidelity
cannot decrease under trace--preserving quantum operation
\cite{barnum}. This property is usually referred to as the
monotonicity under quantum operations \cite{nielsen}. In this
section we show that the partial fidelities possess the same
monotonicity but with respect to unistochastic operations. From
the physical point of view, the most general operation on the
principal system $A$ is to allow $A$ to interact unitarily with
$N$-dimensional environment $E$ in a standard (normalized) state.
The final state of $A$ is then obtained by the operation of
partial trace. Let $\Theta$ denote the initial standard state of
environment. A linear map ${\cal{E}}$ is defined by
\begin{equation}
{\cal{E}}(\rho):={\rm{tr}}_E\left(
{\mathsf{U}}(\rho\otimes\Theta){\mathsf{U}}^{\dagger}\right)
\ , \label{operdef}
\end{equation}
where unitary operator ${\mathsf{U}}$ acts on the space
${\cal{H}}_A\otimes{\cal{H}}_E$ and the trace is taken over space
${\cal{H}}_E$ of environment. This is environmental representation
of the map \cite{bengtsson}. It is commonly to check that the
above map is linear and completely positive. Due to the properties
of the trace and ${\rm{tr}}_E(\Theta)=1$, one also holds
\begin{equation}
{\rm{tr}}_A({\cal{E}}(\rho))={\rm{tr}}_A(\rho)\>{\rm{tr}}_E(\Theta)={\rm{tr}}_A(\rho)
\ . \label{oper2}
\end{equation}
So the map ${\cal{E}}$ is trace--preserving. For all the normalized
inputs $\rho$, we have ${\rm{tr}}_A({\cal{E}}(\rho))=1$. This
gives the probability that the described physical process occurs.
Thus, the considered process is deterministic. Trace--preserving
completely positive maps are known under various names
\cite{bengtsson}: deterministic quantum operations, quantum
channels, stochastic maps. The important case of Eq.
(\ref{operdef}) is given, when the initial state of the
environment is maximally mixed, $\Theta={\mathbf{1}}_E/N$. Such
quantum channels are called {\it unistochastic} \cite{bengtsson}.
Their physical reason is natural. With no knowledge about the
environment, one assumes that it is initially in the maximally
mixed state. Unistochastic quantum operation is analog of
classical transformations given by unistochastic matrices
\cite{bengtsson}. It is clear that unistochastic channels leave
the maximally mixed state ${\mathbf{1}}_A/d$ invariant. So they
are unital maps. In the one-qubit case, both the depolarizing and
phase damping channels are unistochastic.

Like the fidelity \cite{barnum}, the monotonicity of partial
fidelities is based on their behaviour under the operation of
partial trace. Let $\rho_A$ and $\omega_A$ be density operators
obtained as traces
\begin{eqnarray}
\rho_A={\rm{tr}}_E(\widetilde{\rho})
\ , \label{rhotil} \\
\omega_A={\rm{tr}}_E(\widetilde{\omega})
\ , \label{ometil}
\end{eqnarray}
from density operators $\widetilde{\rho}$ and $\widetilde{\omega}$ on the space ${\cal{H}}_A\otimes{\cal{H}}_E$.
To find the necessary property, we consider the quantity
\begin{equation}
G_k(\rho_A,\omega_A|{\mathsf{X}},{\mathsf{Y}})={\rm{tr}}_A(\rho_A{\mathsf{X}})+{\rm{tr}}_A(\omega_A{\mathsf{Y}})
\ , \label{gxy}
\end{equation}
where positive operators ${\mathsf{X}}$ and ${\mathsf{Y}}$ obey
Eqs. (\ref{xpair}) and (\ref{ypair}),
${\rm{rank}}({\mathsf{X}})={\rm{rank}}({\mathsf{Y}})=d-k$. We now
put the operators
\begin{eqnarray}
 & \widetilde{\mathsf{X}}:={\mathsf{X}}\otimes{\mathbf{1}}_E
\ , \label{xwid} \\
 & \widetilde{\mathsf{Y}}:={\mathsf{Y}}\otimes{\mathbf{1}}_E
\ , \label{ywid}
\end{eqnarray}
with the identity ${\mathbf{1}}_E$ on the environment space. They satisfy
$\widetilde{\mathsf{X}}\widetilde{\mathsf{Y}}\widetilde{\mathsf{X}}=\widetilde{\mathsf{X}}$ and
$\widetilde{\mathsf{Y}}\widetilde{\mathsf{X}}\widetilde{\mathsf{Y}}=\widetilde{\mathsf{Y}}$.
Due to the properties of partial trace \cite{nielsen,watrous1}, we have
\begin{eqnarray}
 & {\rm{tr}}\bigl(\widetilde{\rho}\>({\mathsf{X}}\otimes{\mathbf{1}}_E)\bigr)
={\rm{tr}}_A\bigl({\rm{tr}}_E(\widetilde{\rho})\>{\mathsf{X}}\bigr)
\ , \label{xpart} \\
 &  {\rm{tr}}\bigl(\widetilde{\omega}\>({\mathsf{Y}}\otimes{\mathbf{1}}_E)\bigr)
={\rm{tr}}_A\bigl({\rm{tr}}_E(\widetilde{\omega})\>{\mathsf{Y}}\bigr)
\ . \label{ypart}
\end{eqnarray}
Therefore, the quantity $G_k(\rho_A,\omega_A|{\mathsf{X}},{\mathsf{Y}})$ is equal to the quantity
\begin{equation}
G_{(kN)}(\widetilde{\rho},\widetilde{\omega}|\widetilde{\mathsf{X}},\widetilde{\mathsf{Y}})=
{\rm{tr}}(\widetilde{\rho}\>\widetilde{\mathsf{X}})+{\rm{tr}}(\widetilde{\omega}\>\widetilde{\mathsf{Y}})
\ . \label{wgxy}
\end{equation}
Hence we obtain the equality
\begin{equation}
\inf\{G_k(\rho_A,\omega_A|{\mathsf{X}},{\mathsf{Y}})\}=
\inf\{G_{(kN)}(\widetilde{\rho},\widetilde{\omega}|\widetilde{\mathsf{X}},\widetilde{\mathsf{Y}}):
\>\widetilde{\mathsf{X}}={\mathsf{X}}\otimes{\mathbf{1}}_E,\>
\widetilde{\mathsf{Y}}={\mathsf{Y}}\otimes{\mathbf{1}}_E\}
\ . \label{qewq}
\end{equation}
We also note that
${\rm{rank}}(\widetilde{\mathsf{X}})={\rm{rank}}(\widetilde{\mathsf{Y}})=(d-k)N$.
So we can apply the Uhlmann result (\ref{uhlres}) to both the
sides of Eq. (\ref{qewq}). The left-hand side of Eq. (\ref{qewq})
is doubled $k$-fidelity $F_k(\rho_A,\omega_A)$. The right-hand
side of Eq. (\ref{qewq}) is conditional infimum which is larger
than or equal to
$2\>F_{(kN)}(\widetilde{\rho},\widetilde{\omega})$. That is, we
have proved
\begin{equation}
F_k\bigl({\rm{tr}}_E(\widetilde{\rho}),{\rm{tr}}_E(\widetilde{\omega})\bigr)
\geq F_{(kN)}(\widetilde{\rho},\widetilde{\omega})
\ . \label{geqpart}
\end{equation}
For $k=0$ we have the well-known result that the fidelity cannot
decrease under operation of partial trace \cite{barnum}. This
provides the intuitive reason that objects become less
distinguishable when only partial information is available
\cite{nielsen}. In general, the partial fidelities enjoy less
property. But we are able to establish the following result.

{\bf Theorem 5.} {\it If the map ${\cal{E}}$ is unistochastic then
for all $k=0,1,\ldots,d-1$ and any two inputs $\rho$ and $\omega$
there holds}
\begin{equation}
F_k({\cal{E}}(\rho),{\cal{E}}(\omega))\geq F_k(\rho,\omega)
\ . \label{monot}
\end{equation}

{\bf Proof.} Each unistochastic operation can be written in the form (\ref{operdef}) with $\Theta={\mathbf{1}}_E/N$.
Using the multiplicative properties, we see that singular values
\begin{equation}
s_{ji}(\sqrt{\rho\otimes\Theta}\sqrt{\omega\otimes\Theta})=
s_j(\sqrt{\rho}\sqrt{\omega})s_i(\Theta)=\frac{1}{N}\>s_j(\sqrt{\rho}\sqrt{\omega})
\ . \label{sijn}
\end{equation}
Hence the sum of $(d-k)N$ smallest numbers $s_{ji}$ is equal to
the sum of $(d-k)$ smallest numbers
$s_{j}(\sqrt{\rho}\sqrt{\omega})$, that is
\begin{equation}
F_{(kN)}(\rho\otimes\Theta,\omega\otimes\Theta)=F_k(\rho,\omega)
\ . \label{fkfkn}
\end{equation}
Due to the unitary invariance, we have
\begin{equation}
F_{(kN)}(\rho\otimes\Theta,\omega\otimes\Theta)=
F_{(kN)}\left({\mathsf{U}}(\rho\otimes\Theta){\mathsf{U}}^{\dagger},
{\mathsf{U}}(\omega\otimes\Theta){\mathsf{U}}^{\dagger}\right)
\ . \label{monot1}
\end{equation}
By partial trace operation and Eq. (\ref{geqpart}), we obtain Eq. (\ref{monot}). $\blacksquare$

Note that the choice $\Theta={\mathbf{1}}_E/N$ is necessary for
validity of Eq. (\ref{fkfkn}) for all inputs. Indeed, there is no
reasons to think that partial fidelities enjoy the whole of
properties of the fidelity. We now show that the monotonicity of
partial fidelities is violated under the amplitude damping with
$\gamma\in(0;1)$. In the Bloch representation, the effect of
amplitude damping is expressed as \cite{nielsen}
\begin{equation}
(u_x,u_y,u_z)\longmapsto
\left(u_x\sqrt{1-\gamma},u_y\sqrt{1-\gamma},\gamma+u_z(1-\gamma)\right)
\ . \label{adbloch}
\end{equation}
Here $\vec{u}$ denotes the Bloch vector of input density matrix.
Choose two inputs $\rho=(1/2)\{{\mathbf{1}}+v\sigma_z\}$ and
$\omega=(1/2)\{{\mathbf{1}}+w\sigma_z\}$, where $\sigma_z$ is the
Pauli matrix and $v,w\in(0;1)$. By calculations, the smallest
eigenvalue of $|\sqrt{\rho}\sqrt{\omega}|$ is equal to
$\sqrt{(1-v)(1-w)}/2=F_1(\rho,\omega)$. As a result of amplitude
damping, the outputs $\rho'=(1/2)\{{\mathbf{1}}+v'\sigma_z\}$ and
$\omega'=(1/2)\{{\mathbf{1}}+w'\sigma_z\}$ have the Bloch vectors
$(0,0,v')$ and $(0,0,w')$ respectively, where
\begin{eqnarray}
 & v'=\gamma+v(1-\gamma)
\ , \label{vnew}\\
 & w'=\gamma+w(1-\gamma)
\ , \label{wnew}
\end{eqnarray}
in line with Eq. (\ref{adbloch}). By analogy, we get
$F_1(\rho',\omega')=\sqrt{(1-v')(1-w')}/2$. It is easy to check
that (except $v=1$ and $w=1$)
\begin{equation}
(1-v')(1-w')<(1-v)(1-w)
\end{equation}
and, therefore, $F_1(\rho',\omega')<F_1(\rho,\omega)$. The
violation of Eq. (\ref{monot}) is natural because the amplitude
damping channel is not unistochastic. Meanwhile, partial fidelity
$F_1$ can be increased under action of amplitude damping. The
maximally mixed state $\rho_{*}={\mathbf{1}}/2$ is transformed
into $\rho''=(1/2)\{{\mathbf{1}}+\gamma\sigma_z\}$. We also take
state $\omega_{*}=(1/2)\{{\mathbf{1}}-\alpha\sigma_z\}$ with
$\alpha=\gamma/(1-\gamma)$ so that it is mapped into the maximally
mixed state, i.e. $\omega''={\mathbf{1}}/2$. To provide
$\alpha\leq1$, we assume that $\gamma\leq1/2$. By simple
calculations, we get $F_1(\rho_{*},\omega_{*})=\sqrt{1-\alpha}/2$
and $F_1(\rho'',\omega'')=\sqrt{1-\gamma}/2$. It is clear that
$\gamma<\alpha$ and
$F_1(\rho'',\omega'')>F_1(\rho_{*},\omega_{*})$ as claimed. Thus,
the monotonicity of partial fidelities is violated under the
action of some stochastic maps. As it is shown in Ref.
\cite{rast091}, the partitioned trace distances cannot increase
under quantum operations of certain kind including bistochastic
maps. Similarly, the partial fidelities are monotone with respect
to subclass of trace--preserving quantum operations.

\section{Conclusion}

We have analyzed some important properties of Uhlmann's partial
fidelities. The equality of the standard fidelity for two pairs of
density operators does not imply their unitary equivalence. It is
for this reason that the partial fidelities were introduced by
Uhlmann. We have obtained simple bounds on $k$-th partial
fidelity. Quantum--classical relations are considered in terms of
quantum measurement. Like usual fidelity, this gives a statistical
interpretation for partial fidelity. The relationship with
partitioned trace distances is obtained. Our reasons are
significantly based on the Ky Fan maximum principle and its
consequences. In a certain sense, the partial fidelities are
sub--multiplicative. It is shown that any partial fidelity cannot
decrease under unistochastic quantum operation. That is, it enjoys
monotonicity. In general, however, the partial fidelities are not
monotone. The derived properties allow to lighten work with the
partial fidelities.


\begin{thebibliography}{77}

\bibitem{uhlmann76}
A.~Uhlmann (1976), {\it The transition probability in the state space of a *--algebra},
Rep. Math. Phys. {\bf 9}, pp. 273--279.

\bibitem{uhlmann83}
P.~M.~Alberti and A.~Uhlmann (1983), {\it Stochastic linear maps and transition probability}, Lett. Math. Phys. {\bf 7}, pp. 107–-112.

\bibitem{jozsa94}
R.~Jozsa (1994), {\it Fidelity for mixed quantum states}, J. Mod. Opt. {\bf 41}, pp. 2315--2323.

\bibitem{nielsen}
M.~A.~Nielsen and I.~L.~Chuang (2000), {\it Quantum computation and
quantum information}, Cambridge University Press (Cambridge).

\bibitem{nielsen05}
A.~Gilchrist, N.~K.~Langford and M.~A.~Nielsen (2005), {\it Distance measures to compare real and ideal quantum processes}, Phys. Rev. A {\bf 71}, 062310.

\bibitem{braunstein}
S.~L.~Braunstein and C.~M.~Caves (1995), {\it Geometry of quantum states}, Proceedings of the Conference on Quantum Communication and Measurement, edited by R.~Hudson, V.~P.~Belavkin, O.~Hirota, New York (Plenum Press), pp. 21--30.

\bibitem{bengtsson}
I.~Bengtsson and K.~\.{Z}yczkowski (2006), {\it Geometry of quantum states: an
introduction to quantum entanglement}, Cambridge University Press (Cambridge).

\bibitem{holevo}
V.~Giovannetti and A.~S.~Holevo (2008), {\it Quantum Shared Broadcasting},
Quantum Information Processing {\bf 7}, pp. 55--69.

\bibitem{nielsen08}
M.~R.~Dowling and M.~A.~Nielsen (2008), {\it The geometry of quantum computation},
Quantum Information {\&} Computation {\bf 8}, pp. 861--899.

\bibitem{fuchs}
C.~A.~Fuchs (1996), {\it Distinguishability and accessible information in
quantum theory}, quant-ph/9601020.

\bibitem{uhlmann00}
A.~Uhlmann (2000), {\it On "partial" fidelities}, Rep. Math. Phys. {\bf 45}, pp. 407--418.

\bibitem{foster}
P.~E.~M.~F.~Mendon\c{c}a, R.~J.~Napolitano, M.~A.~Marchiolli, C.~J.~Foster
and Y.--C.~Liang (2008), {\it Alternative fidelity measure between quantum states},
Phys. Rev. A {\bf 78}, 052330.

\bibitem{uhlmann09}
J.~A.~Miszczak, Z.~Pucha{\l}a, P.~Horodecki, A.~Uhlmann and K.~\.{Z}yczkowski (2009),
{\it Sub–- and super–-fidelity as bounds for quantum fidelity},
Quantum Information {\&} Computation {\bf 9}, pp. 0103--0130.

\bibitem{rast031}
A.~E.~Rastegin (2003), {\it Upper bound on the global fidelity for mixed--state cloning},
Phys. Rev. A {\bf 68}, 012305.

\bibitem{kzws01}
K.~\.{Z}yczkowski and W.~S{\l}omczy\'{n}ski (1998), {\it The Monge
distance between quantum states}, J. Phys. A: Math. Gen. {\bf 31},
pp. 9095--9104.

\bibitem{rast06}
A.~E.~Rastegin (2006), {\it Sine distance for quantum states}, quant-ph/0602112.

\bibitem{watrous1}
J.~Watrous (2008), {\it CS 798: Theory of quantum information}, University of Waterloo,
http://www.cs.uwaterloo.ca/$\sim$watrous/quant--info/lecture--notes/all--lectures.pdf

\bibitem{bhatia}
R.~Bhatia (1997), {\it Matrix Analysis}, Springer (New York).

\bibitem{caves}
C.~A.~Fuchs and C.~M.~Caves (1995), {\it Mathematical techniques for quantum communication
theory}, Open Systems {\&} Information Dynamics {\bf 3}, pp. 345--356.

\bibitem{rast032}
A.~E.~Rastegin (2003), {\it Global--fidelity limits of state--dependent cloning of mixed states},
Phys. Rev. A {\bf 68}, 032303.

\bibitem{alulm00}
P.~M.~Alberti and A.~Uhlmann (2000), {\it On Bures distance and *--algebraic transition
probability between inner derived positive linear forms over W*--algebras},
Acta Applicandae Mathematicae {\bf 60}, pp. 1--37.

\bibitem{peres}
A.~Peres (1993), {\it Quantum theory: concepts and methods}, Kluwer (Dordrecht).

\bibitem{kyfan}
K.~Fan (1949), {\it On a theorem of Weyl concerning eigenvalues of linear transformations. I},
Proc. Nat. Acad. Sci. U.S.A. {\bf 35}, pp. 652--655.

\bibitem{vidal}
M.~A.~Nielsen and G.~Vidal (2001), {\it Majorization and the interconversion of bipartite states},
Quantum Information {\&} Computation {\bf 1}, pp. 76--93.

\bibitem{barnum}
H.~Barnum, C.~M.~Caves, C.~A.~Fuchs, R.~Jozsa and B.~Schumacher (1996), {\it Noncommuting mixed
states cannot be broadcast}, Phys. Rev. Lett. {\bf 76}, pp. 2818--2821.

\bibitem{puchala}
Z.~Pucha{\l}a and J.~A.~Miszczak (2009), {\it Bound on trace distance based on super--fidelity},
Phys. Rev. A {\bf 76}, 024302.

\bibitem{rast091}
A.~E.~Rastegin (2009), {\it Partitioned trace distances}, arXiv:0903.4543 [quant-ph].

\bibitem{graaf}
C.~A.~Fuchs and J. van de Graaf (1999), {\it Cryptographic distinguishability measures
for quantum mechanical states}, IEEE Trans. Inf. Theory {\bf 45}, pp. 1216--1227.

\bibitem{rast07}
A.~E.~Rastegin (2007), {\it Trace distance from the viewpoint of quantum
operation techniques}, J. Phys. A: Math. Theor. {\bf 40}, 9533--9549.

\end{thebibliography}
\end{document}